\documentclass[aps,twocolumn,epsfig,superscriptaddress,prb]{revtex4}

\usepackage{amsmath}
\usepackage{graphicx}
\usepackage{psfrag}

\begin{document}

\title{Controlling edge states of zigzag carbon nanotubes by the
Aharonov-Bohm flux} 

\author{K. Sasaki}
\email[Email address: ]{sasaken@imr.tohoku.ac.jp}
\affiliation{Institute for Materials Research, Tohoku University, 
Sendai 980-8577, Japan}

\author{S. Murakami}
\affiliation{Department of Applied Physics, University of Tokyo, Hongo,
Bunkyo-ku, Tokyo 113-8656, Japan}

\author{R. Saito}
\affiliation{Department of Physics, Tohoku University and CREST, JST,
Sendai 980-8578, Japan}

\author{Y. Kawazoe}
\affiliation{Institute for Materials Research, Tohoku University, 
Sendai 980-8577, Japan}

\date{\today}
 
\begin{abstract}
 It has been known theoretically that localized states exist around
 zigzag edges of a graphite ribbon and of a carbon nanotube, whose
 energy eigenvalues are located between conduction and valence bands. 
 We found that in metallic single-walled zigzag carbon 
 nanotubes two of the localized states become critical, and that their
 localization length is sensitive to the mean curvature
 of a tube and can be controlled by the Aharonov-Bohm flux.
 The curvature induced mini-gap closes by the relatively weak magnetic
 field. 
 Conductance measurement in the presence of the Aharonov-Bohm flux can
 give information about the curvature effect and the critical states.
\end{abstract}

\pacs{}
\maketitle

Prior theoretical studies on zigzag carbon nanotubes clarified that they
can exhibit either metallic or semiconducting energy band depending on
their chiral vector~\cite{SDD}.
It has been shown theoretically~\cite{Saito} and
experimentally~\cite{Ouyang} that even in
``metallic'' zigzag single-walled carbon nanotubes (SWCNTs) 
a finite curvature opens small energy gaps. 
Therefore, any zigzag SWCNTs have finite energy gaps.

On the other hand, Fujita {\it et al}.~\cite{Fujita} theoretically
showed that localized states (edge states) emerge at graphite zigzag
edges. 
The edge states are plane wave modes along the edge and their energy
eigenvalues are in between the valence band and the conduction band
(zero-energy states).
Since the graphite sheet with zigzag edges can be rolled to form a
zigzag SWCNT, the edge states are supposed to be localized at both edges
of the zigzag SWCNT.
In this case, a zigzag nanotube has not only bulk (extended) states with
a finite energy gap but also zero-energy localized edge states.
Although several properties of the edge state have been
investigated~\cite{Fujita2}, physical relationship between electrical
properties of bulk states and edge states remains to be
clarified. 

\begin{figure}[htbp]
 \begin{center}
  \psfrag{L}{Left edge}
  \psfrag{R}{Right edge}
  \psfrag{a}{$\psi_{AI}^0$}
  \psfrag{b}{$\psi_{AI+1}^0$}
  \psfrag{c}{$\psi_{AI'}^1$}
  \psfrag{d}{$\psi_{BI'}^{N-1}$}
  \psfrag{e}{$\psi_{BI}^N$}
  \psfrag{f}{$\psi_{BI+1}^N$}
  \psfrag{1}{0}
  \psfrag{2}{1}
  \psfrag{3}{2}
  \psfrag{N}{$N$}
  \psfrag{i}{$I$}
  \psfrag{j}{$J$}
  \psfrag{g}{$\Phi$}
  \includegraphics[scale=0.3]{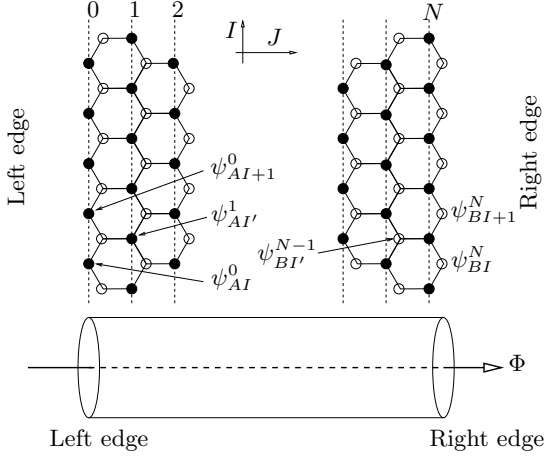}
 \end{center}
 \caption{Lattice structure of a zigzag carbon nanotube. 
 The filled (open) circle indicates the A (B) sublattice.
 Both the left and right edges are Fujita's edges.
 }
 \label{fig:edge}
\end{figure}

In this paper, to investigate this relationship, we study an effect of
the Aharonov-Bohm (AB) flux along the metallic zigzag nanotube axis.
Among the Fujita's edge states of metallic zigzag SWCNTs, we will show
that there exist ``critical states''.
Their wave functions, in particular their localization length, are
sensitive to the following two perturbations: the curvature and the
Aharonov-Bohm (AB) flux. 
These perturbations are new ingredients for cylindrical geometry, absent
in the flat graphene sheet. 
The main purpose of the paper is to clarify the dependence of the
critical states on the AB flux and the relationship between the wave
functions at the bulk and at the edge.
This dependence can be examined by a conductance measurement in the
presence of the AB flux. 
We note that such AB flux applied along the SWCNT axis (see
Fig.~\ref{fig:edge}) has already been realized in
experiments~\cite{Zaric,Minot}.

Because the unit cell is composed of two sublattices, A and B,
we write the wave functions as
$| \Psi_{k} \rangle = {}^t\left( \psi_A(k),\psi_B(k) \right)$
where $k$ is a discrete wave vector around the tubule axis.
By fixing the chiral vector~\cite{SDD} as $C_h = (n,0)$, we obtain
$k = 2\pi \mu/|C_h|$ $(|C_h|=\sqrt{3}a_{\rm cc} n)$ where
$\mu$ $(=1,\cdots,n)$ is an integer and $a_{\rm cc} \approx$ 1.42\AA\ is
the carbon-carbon bond length.
We analyze the system using the nearest-neighbor tight-binding
Hamiltonian,
${\cal H} = \sum_{a=1,2,3} \sum_{i \in A} \left( V_\pi + \delta V_a
\right) a_{i+a}^\dagger a_i + h.c.$. 
``A'' (in the summation index) denotes an A-sublattice, $a_i$ and
$a_i^\dagger$ are canonical annihilation-creation operators of the 
electron at site $i$, and site $i+a$ indicates the nearest-neighbor
sites ($a = 1,2,3$) of site $i$. 
We include curvature effect as the bond direction-dependent hopping
integral, $V_\pi + \delta V_a$.
We ignore the electron spin for simplicity.

The energy eigen equation, 
${\cal H} |\Psi_{k} \rangle = E | \Psi_{k} \rangle$, becomes
\begin{align}
 & \epsilon \phi_A^{J+1} = \phi_B^J + g \phi_B^{J+1} \ \
 (J=0,\cdots,N-1), 
 \label{eq:sq-1} \\
 & \epsilon \phi_{B}^J = \phi_A^{J+1} + g \phi_A^J \ \
 (J=0,\cdots,N-1), 
 \label{eq:sq-2} \\
 & \epsilon \phi_A^0 = g \phi_B^0, \ \ \
 \epsilon \phi_B^N = g \phi_A^N,
 \label{eq:bound-con}
\end{align}
where we define
\begin{align}
 \epsilon \equiv \frac{E}{(V_\pi + \delta V_1)},\ \
 g \equiv 2 \frac{(V_\pi + \delta V_2)}{(V_\pi + \delta V_1)}
 \cos \frac{\pi(\mu - n_\Phi)}{n}.
 \label{eq:g-value}
\end{align}
Here, we write the wave function at site $(I,J)$ of A(B) sublattice as  
$\psi_{(A,B)I}^J(k) = \exp(i 2\pi I \mu/n) \phi_{(A,B)}^J(\mu)$ 
(see Fig.~\ref{fig:edge}).
We assume $\delta V_2 = \delta V_3$ in the Hamiltonian because of the
mirror symmetry along the axis of a zigzag nanotube. 
The effect of the AB flux is included in $g$ of Eq.~(\ref{eq:g-value})
by the replacement, $\mu \to \mu - n_\Phi$, where 
$n_\Phi \equiv \Phi/\Phi_0$ is the number of flux quantum, $\Phi_0$.

By solving Eqs.~(\ref{eq:sq-1}), (\ref{eq:sq-2}) and
(\ref{eq:bound-con}), we obtain an analytical form of the energy
eigen function as 
\begin{align}
 & \phi_A^J =  \left( \frac{1}{g} \frac{\sin J \phi}{\sin \phi} + 
 \frac{\sin (J+1)\phi}{\sin \phi} \right) \phi_A^0, 
 \label{eq:sol-ex-A} \\
 & \phi_B^J = \left(
 \frac{\sin (J+1) \phi}{\sin \phi} \right) \phi_B^0
 \ \ (J=0,\cdots,N),
 \label{eq:sol-ex-B}
\end{align}
where $\phi$ satisfies 
\begin{align}
 2 \cos \phi =  \frac{\epsilon^2 - g^2 -1}{g} \equiv \kappa.
 \label{eq:def-kappa}
\end{align}
The energy eigenvalue is determined by the boundary condition of 
Eq.~(\ref{eq:bound-con}). 
Using Eqs.~(\ref{eq:sol-ex-A}) and (\ref{eq:sol-ex-B}), we get 
\begin{align}
 \frac{\sin (N+1)\phi + g \sin (N+2) \phi}{\sin \phi} = 0.
 \label{eq:bound-con-ex}
\end{align}
This equation corresponds to vanishing wave function at fictitious A
sites of $J=N+1$, i.e., $\phi_A^{N+1} = 0$. 
Most of the solutions for $\phi$ in Eq.~(\ref{eq:bound-con-ex}) are
real, as we explain later. Such real solutions represent extended states
and satisfy $\kappa^2\le 4$.

In addition, there can be localized states, where $\phi$ has an
imaginary part and $\kappa^2>4$. 
Their localization length is proportional to the inverse of the
imaginary part of $\phi$.
We examine if the boundary condition allows such a localized state.
The complex solutions of Eq.~(\ref{eq:bound-con-ex}) can be written as
$\phi = i\varphi$ or $\phi = \pi + i\varphi$ where $\varphi$ is a real
number.
The former case, $\phi=i\varphi$, corresponds to $\kappa > 2$.
In this case, Eq.~(\ref{eq:bound-con-ex}) becomes
\begin{align}
 \sinh (N+1) \varphi + g \sinh (N+2)\varphi = 0.
\end{align}
Due to $|\sinh (N+2)\varphi| > |\sinh (N+1) \varphi|$, we obtain
$-1< g <0$. 
The latter case, $\phi=\pi+i\varphi$, corresponds to $\kappa < -2$ and
$0< g<1$. 
When $|g|\geq 1$, on the other hand, Eq.~(\ref{eq:bound-con-ex}) does
not allow complex solutions and therefore there is no localized state in
this region. 
Because the boundary condition implies
$g = - e^{+\varphi} + {\cal O}(e^{2N\varphi})$ for
$\varphi < 0 \ (-1 < g < 0)$ and 
$g = e^{-\varphi} + {\cal O}(e^{-2N\varphi})$ for 
$\varphi > 0 \ (0 < g < 1)$, 
the energy eigenvalues for localized states are 
$\epsilon = \pm {\cal O}(e^{-N|\varphi|})$,
exponentially small as a function of nanotube length.

By a more elaborated analysis, for a fixed $k$,  we can analytically
show that, 
(i) for $|g|\geq\frac{N+1}{N+2}$, all the $2(N+1)$ states are extended,
(ii) for $0<g<\frac{N+1}{N+2}$, there are $2N$ extended states and two
localized states with ${\rm Re}\phi=\pi$, and  
(iii) for $-\frac{N+1}{N+2}<g<0$, there are $2N$ extended states and two
localized states with ${\rm Re}\phi=0$.
For each wave vector $k$ satisfying $|g|<\frac{N+1}{N+2}$, the two
localized states have energies with opposite signs, 
$\epsilon=\pm {\cal O}(e^{-N|\varphi|})$.
Each of the two states is localized near both two edges. 
In the left (right) edge, it is localized in the A (B) sublattice.
Henceforth we consider the length $N$ of the nanotube to be 
large; the localized states are then allowed for $|g|<1$.

\begin{figure}[htbp]
 \begin{center}
  \psfrag{a}{(a)}
  \psfrag{b}{(b)}
  \psfrag{e}{$\epsilon$}
  \psfrag{g}{$g$}
  \psfrag{1}{$1$}
  \psfrag{-1}{$-1$}
  \psfrag{2}{$2$}
  \psfrag{-2}{$-2$} 
  \psfrag{3}{$3$}
  \psfrag{-3}{$-3$}
  \psfrag{d}{$g=-\left( 1 + \frac{\delta V_2 - \delta V_1}{V_\pi} \right)$} 
  \includegraphics[scale=0.3]{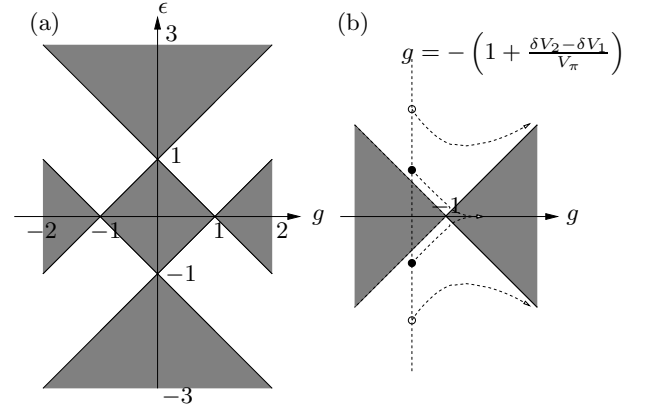}
 \end{center}
 \caption{(a) Region for localized states $\kappa^{2}>4$ shown as shaded
 area in the $(g,\epsilon)$ plane.
 Whether localized states are allowed depends on the boundary
 condition. 
 If the figure is rewritten in terms of $k$ and $E$, the empty region
reproduces 
 the well-known band structure for the graphene sheet. 
 (b) When we ignore the curvature effect, there are states with 
 $g=-1$ in the absence of the AB flux.
 The curvature effect displaces the states onto
 the line  $g=-(1+(\delta V_2-\delta V_1)/V_\pi)$. 
 The AB flux changes the value of $g$, and eigenstates will 
 make trajectories like the dashed curves in the figure.
 Two extended states (filled circles) become 
 localized as they run into the shadow region. 
 Eigenstates except for these two states remain extended
 (open circles).
} 
 \label{fig:critical}
\end{figure}

The critical condition, $\kappa^2 = 4$, separates the extended and the
localized states. 
We plot the lines of the critical condition in the $(g,\epsilon)$ plane
in Fig.~\ref{fig:critical}(a).
The shadow regions satisfy $\kappa^2 > 4$, representing localized states.
By applying the AB flux, each state moves and makes a trajectory in the
$(g,\epsilon)$ plane. 
Suppose one extended state, located outside of the shaded region, comes
across the boundaries $\kappa^2=4$ between the empty and shaded
regions. 
It means that the extended state turns into an localized state.
On the verge of the transition the state becomes ``critical'',
when $g=\pm \frac{N+1}{N+2}\approx \pm 1$ and 
$\epsilon = \pm \frac{1}{N+2}$. 
If we assume $\delta V_a=0$ and there is no external magnetic field,
this condition for $g$ is satisfied only in metallic zigzag nanotubes,
namely, when $n$ is a multiple of 3 and $\mu_1/n=1/3$ or $2/3$ (see
Eq.~(\ref{eq:g-value})). 
When it is satisfied, the states with $g=\pm 1$, $\epsilon\approx 0$ are
located very close to the critical line $\kappa^{2}=4$, and thus can be
easily controlled by external perturbations as we see later.

The cylindrical geometry of nanotubes yields a finite mean curvature and
induces a change of the hopping integral $\delta V_a$~\cite{Saito,KM}.
The scaling of the curvature gives
$\delta V_a/V_\pi \approx {\cal O}(a_{\rm cc}^2/|C_h|^2)$.
The values of $g$ are then driven away from $g=\pm 1$ to 
\begin{align}
 g \approx \pm \left( 1 + \frac{\delta V_2-\delta V_1}{V_\pi} \right)
 + \sqrt{3} \frac{\pi}{n} n_\Phi.
 \label{eq:app-r}
\end{align}
From an experimental data of the mini-gap in metallic zigzag
nanotubes~\cite{Ouyang}, we estimate~\cite{SKS}
$\delta V_2 - \delta V_1 = \pi^2 V_\pi/8n^2$.
Thus, in the absence of the AB flux, these states have
$|g| > 1$, and are not critical but extended.
However, the AB flux can make one of $g$ within $|g| < 1$,
which implies that two extended states become localized
(see Fig.~\ref{fig:critical}(b)).
Other extended states remain extended even with this external magnetic
field. 

One of the important questions is whether the AB flux required for 
such a transition is experimentally accessible or not.
We calculate the magnetic field required to close the curvature-induced
mini-gap. 
From Eq.~(\ref{eq:app-r}), we find $\bar{n}_\Phi = \pi/(8\sqrt{3}n)$ 
is necessary to close the mini-gap for $g=-1$
(for $g=1$, $-\bar{n}_\Phi$ is necessary).
This AB flux corresponds to the magnetic field $
B \approx \frac{2 \times 10^5}{n^3}[{\rm T}]$.
Thus, $B$ is accessible if $n > 20$.
For a zigzag nanotube with $C_h=(n,0)$, the flux quantum 
$\Phi_0$ corresponds to the magnetic field 
$B_n = B_1/n^2 [{\rm T}]$ 
($B_1 \equiv 8.5\times 10^5[{\rm T}]$), and therefore 
experimentally accessible magnetic flux corresponds to $|n_\Phi| \ll 1$.
For instance, we have $B_9 \approx 10^4 [{\rm T}]$, which is well beyond
an accessible magnetic field $\approx10^2[{\rm T}]$, and 
we can attain $|n_\Phi| \approx 10^{-2}$ at most.

\begin{figure}[htbp]
 \begin{center}
  \psfrag{t}{$L/|C_h|$}
  \psfrag{a}{(a) $n=30$}
  \psfrag{b}{(b) $n=51$}
  \psfrag{B}{$B[{\rm T}]$}
  \psfrag{R1}{$|C_h|=7.3{\rm nm}$}
  \psfrag{R2}{$|C_h|=12.5{\rm nm}$}
  \includegraphics[scale=0.35]{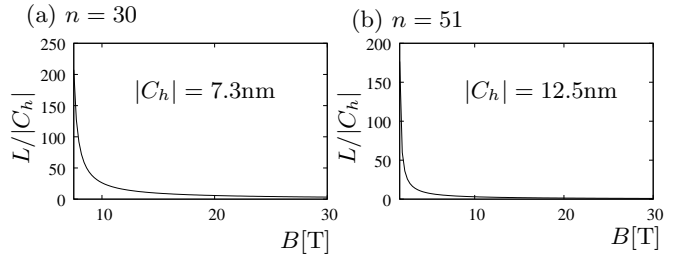}
 \end{center}
 \caption{Localization length of the critical state for (a) $n=30$ and
 (b) $n=51$ zigzag nanotubes.
 The horizontal axis is a magnetic field, $B[{\rm T}]$, 
 (a) $7.5 \le B \le 30$ and (b) $1.6 \le B \le 30$.}
 \label{fig:llength}
\end{figure}

Here, we plot the localization length of the critical state for metallic
zigzag nanotubes with different diameters.
We define the localization length $L$ by
$|\phi_A^J|^2/|\phi_A^0|^2 \approx \exp(- 3a_{\rm cc} J/2L)$,
where $3a_{\rm cc}/2$ is the distance between $J$ and $J+1$ lines 
(Fig.~\ref{fig:edge}).
We obtain for $n_\Phi > \bar{n}_\Phi$
\begin{align}
 \frac{L}{|C_h|} = \frac{1}{4\pi n_\Phi- \frac{\pi^2}{2\sqrt{3} n}}
 = \frac{1}{\frac{4\pi B}{B_n} -
 \frac{\pi^2}{2\sqrt{3}n}}. 
\end{align}
In Fig.~\ref{fig:llength}, we plot $L/|C_h|$ for $n=30$ taken as the
largest diameter for a SWCNT and $n=51$ taken as a shell in a
multi-walled carbon nanotube (MWCNT).
The curvature-induced energy gaps are $E_{\rm gap}=7.4[{\rm meV}]$ and
$2.6 [{\rm meV}]$, respectively, where we use $V_\pi=2.7[{\rm eV}]$.
$\bar{n}_\Phi$ is estimated as $7.1[{\rm T}]$ for $n=30$ and 
$1.4[{\rm T}]$ for $n=51$.
If we neglect the interlayer interaction in a MWCNT, a zigzag SWCNT in a
MWCNT is the most suitable to examine the behavior of the critical
states, because one does not need strong magnetic field to close the
curvature-induced mini-gap.

\begin{figure}[htbp]
 \begin{center}
  \psfrag{a}{(a) $n_\Phi=0$}
  \psfrag{b}{(b) $0 < n_\Phi < \bar{n}_\Phi$}
  \psfrag{c}{(c) $n_\Phi=\bar{n}_\Phi$}
  \psfrag{d}{(d) $n_\Phi > \bar{n}_\Phi$}
  \psfrag{V}{$V_g$}
  \psfrag{s}{$G_0$}
  \psfrag{t}{$2G_0$}
  \psfrag{cs}{critical states}
  \includegraphics[scale=0.38]{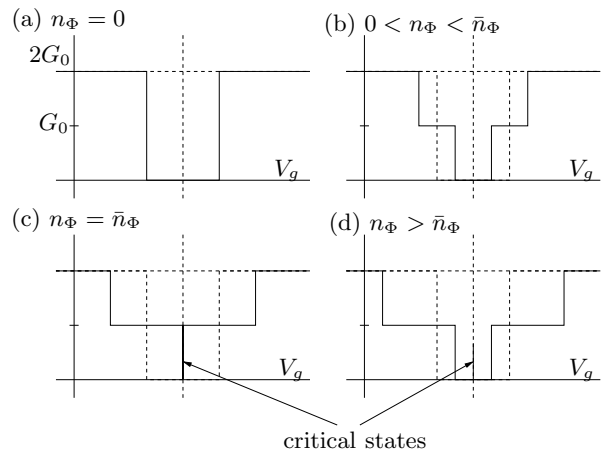}
 \end{center}
 \caption{Theoretical result of conductance with a metallic zigzag
 nanotube.
 We plot conductance as a function of gate voltage, $V_g$ for several
 values of the AB flux.}  
 \label{fig:conduc}
\end{figure}

The curvature effect and the modulation of localization
length can be experimentally proved by 
conductance measurements as a function of the gate
voltage for various values of the AB flux. 
We plot the conductance for different $n_\Phi$ in
Fig.~\ref{fig:conduc}, where we assume ideal contacts and zero
temperature.  
Such measurement is attainable, as 
conductance of 
well-contacted individual SWCNTs was measured in the ballistic regime by
Kong {\it et al}~\cite{Kong}.
In Fig~\ref{fig:conduc}(a), the solid curve represents conductance in the
absence of the AB flux, $n_\Phi = 0$.
The zero conductance state corresponds to the curvature-induced mini-gap.
In (b), conductance changes according to the AB flux. 
Near the Fermi level, only one channel ($g \approx -1$) contributes to
the transport and gives a unit of quantum conductance $G_0$.  
In (c), when $n_{\Phi}=\bar{n}_\Phi$, the zero conductance state
disappears and the critical states start to localize. 
Even when further AB flux is applied, there remains a finite conductance
at Fermi level due to the critical states as is depicted in (d).
However, because further AB flux quickly reduces the transmission
probability from Fig.~\ref{fig:llength}, the conductance at the Fermi
level decreases.

The transition between extended and localized states by the
AB flux can be also understood in terms of the effective-mass theory of
graphene.
The effective-mass theory of low-energy dynamics is given by two Weyl
equations~\cite{SW}, each of which represents the dynamics around the K 
point and K' point in the momentum space.
Curvature in the nanotube 
induces an effective gauge field in the Weyl equations~\cite{KM}, and 
this gauge field gives the mini-gap of $2(\delta V_2-\delta V_1)$ through
the AB effect.
When we apply the AB flux ($n_\Phi < \bar{n}_\Phi$), the energy gap of
one mode opens while that of the other
closes; therefore, the energy gap closes as a total system.
If further magnetic flux ($n_\Phi > \bar{n}_\Phi$) is applied, some of the 
extended states become localized at the edges and their energy becomes zero, 
while other extended states open up the gap again.
Such zero-energy states cannot be obtained by 
solving the Weyl equations with a uniform energy gap.
Instead, it is known that by modulating locally the energy gap,
there appear localized states called Jackiw-Rebbi (JR) states~\cite{JR}.
The edge states in the nanotube induced by the AB flux resemble the JR
states, in that they have zero energy and are localized; meanwhile they
are different from ours because these edge states are induced by a
uniform AB flux, and not by a local modulation. 
We note that related states can appear when we introduce a local
geometrical deformation in the nanotube~\cite{SKS}.

Throughout this paper, we have examined zigzag nanotubes having two 
Fujita's edges.
Even for zigzag nanotubes with open edges, one can consider other cases
with one or two edges being the Klein's edge~\cite{Klein}
If one of the edges is Klein's edge and the other is Fujita's edge, 
there are still localized states, whereas their properties are
distinct from the previous case for two Fujita's edges.
Such a nanotube can be made by attaching A-sites at the right edge of
the zigzag nanotube in Fig.~\ref{fig:edge}.
The boundary condition for the right edge is 
$\epsilon \phi_A^{N+1} = \phi_B^N$.
For localized states we obtain
$\epsilon = 0$ and $\phi_A^J=(-g)^{J}\phi_{A}^0$, $\phi_B^J=0$
($J=0,\cdots,N$). 
Thus, the localized states have exactly $\epsilon=0$ for every value of
$g$, and they have amplitude only on the A sublattice. 
Such states are localized near the right (left) edge when $|g|>1$
($|g|<1$). 
Furthermore, there are no transitions between localized and extended
states, even when we apply a magnetic field.
Meanwhile, when $g$ passes $g=\pm 1$, the localized state with
$\epsilon=0$ comes on the critical boundary of $\kappa^{2}=4$ in
Fig.~\ref{fig:critical}(a), and the corresponding localized state will
move from the right edge to the left or vice versa, and can be detected
experimentally. 
On the other hand, when both edges are the Klein's edges, 
the situation is somewhat similar to the case with two Fujita's edges.
For this case, localized states are realized when $|g| > 1$ instead 
of $|g|<1$, and the AB flux induces a transition from localized to
extended states. 

Similar phenomena are expected also in nanotubes with 
other chiral structures except for armchair nanotubes.
Nakada {\it et al}. showed numerically that localized states
appear not only in the zigzag edges but 
also in edges with other shapes~\cite{Fujita2}.
We expect that such localized states will undergo transition with 
extended states in the presence of the AB flux.

Finally let us mention the Coulomb charging energy for the localized
states.  
A typical energy scale of the charging energy for an edge state is  
$E_c \approx e^2/L_{\rm edge}$, where $L_{\rm edge}$ is the typical
length for the edge states. This energy should be compared with that of
the extended states evaluated as $E_c \approx e^2/L_{\rm sys}$,
where $L_{\rm sys}$ is the system size.
Since $L_{\rm sys} \gg L_{\rm edge}$, the Coulomb energy $E_c$ for the
edge states are much bigger than that of the extended states, and may
hinder localization.
To lower this Coulomb charging energy, the spins of the localized states
at each edge will align ferromagnetically\cite{Fujita}.

In addition to the edge states, the bulk states can also have
interesting phenomena due to the AB flux in SWCNTs.
Namely, the AB flux induces a persistent current around the tube axis, 
and gives an orbital magnetic moment~\cite{Imry}.
The persistent current is caused by a splitting of the van Hove
singularities by the AB flux~\cite{AA}. 
Quite recently, the splitting was observed in semiconducting nanotubes
as the splitting of the first-subband magneto-absorption
peak~\cite{Zaric} and also in small-bandgap (not curvature related)
nanotubes as a temperature dependence of conductance~\cite{Minot}. 
These experiments were intended to observe that the AB flux can make
asymmetry between two energy bands composed of the extended (bulk) wave 
functions near the Fermi level. 
By means of doping in addition to the AB flux, 
an interference between many energy bands~\cite{SMS} takes place.
This interference affects magnetic and transport properties of the
system, and is helpful to get insight into the bulk electronic states.

In summary, the Aharonov-Bohm effect of carbon nanotubes will be suited
to examine not only the bulk electrical properties but also the properties
of the edge states.
We point out that there are two critical states in metallic zigzag
nanotubes.
Although the critical states become extended due to the curvature effect
in the absence of magnetic field, the Aharonov-Bohm flux can make a
transition from the extended states into localized edge states.
This transition can be seen as a characteristic feature of conductance. 

K. S. acknowledges support form the 21st Century COE Program of the
International Center of Research and Education for Materials of Tohoku
University.
S. M. is supported by Grant-in-Aid (No.~16740167) from the Ministry of
Education, Culture, Sports, Science and Technology (MEXT), Japan.
R. S. acknowledges a Grant-in-Aid (Nos.~13440091 and 16076201) from
MEXT.

\end{document}